# The unbalanced reorganization of weaker functional connections induces the altered brain network topology in schizophrenia.

Rossana Mastrandrea[1,*], Fabrizio Piras[2,*], Andrea Gabrielli[3,4], Nerisa Banaj[2], Guido Caldarelli[1,4], Gianfranco Spalletta[2,5,+,**], Tommaso Gili[1,**]

[1]Networks Unit, IMT School for Advanced Studies, Lucca, Italy, 55100
[2]Laboratory of Neuropsychiatry, IRCCS Santa Lucia Foundation, Rome, Italy 00179
[3]Dipartimento di Ingegneria, Università Roma Tre, Rome, Italy 00146
[4]Istituto dei Sistemi Complessi (ISC) - CNR, UoS Sapienza, Dipartimento di Fisica, Università "Sapienza", Rome, Italy 00185
[5]Division of Neuropsychiatry, Menninger Department of Psychiatry and Behavioral Sciences, Baylor College of Medicine, Houston, Tx, USA 77030

*These authors share the First authorship
**These authors share the Senior authorship
+corresponding author: g.spalletta@hsantalucia.it

**Abstract**

Network neuroscience shed some light on the functional and structural modifications occurring to the brain associated with the phenomenology of schizophrenia. In particular, resting-state functional networks have helped our understanding of the illness by highlighting the global and local alterations within the cerebral organization. We investigated the robustness of the brain functional architecture in 44 medicated schizophrenic patients and 40 healthy comparators through an advanced network analysis of resting-state functional magnetic resonance imaging data. The networks in patients showed more resistance to disconnection than in healthy controls, with an evident discrepancy between the two groups in the node degree distribution computed along a percolation process. Despite a substantial similarity of the basal functional organization between the two groups, the expected hierarchy of healthy brains' modular organization is crumbled in schizophrenia, showing a peculiar arrangement of the functional connections, characterized by several topologically equivalent backbones. Thus, the manifold nature of the functional organization's basal scheme, together with its altered hierarchical modularity, may be crucial in the pathogenesis of schizophrenia. This result fits the disconnection hypothesis that describes schizophrenia as a brain disorder characterized by an abnormal functional integration among brain regions.

**Keywords**

FMRI, Resting-State, Schizophrenia, Graph Theory, Percolation, Maximum Spanning Tree, Allometric Scaling

**Data Availability Statement**

The data that support the findings of this study are available from the corresponding author upon reasonable request.

**Conflict of Interest**

The authors declare no conflict of interest.

## Introduction

Data-driven connectivity analysis of resting-state functional magnetic resonance imaging (rs-fMRI)[1,2] has revealed abnormalities in brain network topology in several mental disorders, particularly schizophrenia (SCZ)[3–6]. Often described in terms of circumscribed alterations in neural circuits[7], SCZ has been explained in terms of aberrant interactions among brain regions leading to an integration of distributed brain networks, referred to as ``misconnection'' or ``dysconnection'' syndrome[8,9]. According to this hypothesis, physiological interactions between distributed neuronal ensembles are critical for the effective coordination of information processing and, consequently, the production of coherent action and cognition[10–15].

Synaptic pruning has been hypothesized to underlie the neuropathology of SCZ[16]. In particular, neural networks in the development result from pruning processes, including expansive growth of synapses followed by activity-dependent elimination[17,18]. A dysfunctional synaptic pruning generally implies normal synapses formation, followed by an unbalanced process of elimination[19]. Brain functional connectivity has been proven abnormal in specific brain circuits in SCZ[20–22] and associated with alterations in perception, thoughts, mood, and behavior, typical of this neuropsychiatric illness[23–27].

Most of the whole-brain approaches to the study of brain functional and structural connectedness showed widespread disturbances in the network dynamics[28–34] and alterations of the whole cerebral functional architecture's modular organization in SCZ[32,35–38]. Nonetheless, the alterations in the global functional integration and the brain's local functional connectedness reported in literature need to be clarified. In fact, previous research addressed the local and global brain connectedness disjointly, characterizing either the collective behavior of specific macroscopic networks (e.g. default mode, salience, somatosensory, etc.) or the role of single brain regions as possible hubs in the whole brain web and their coordination with the first neighbours.

In order to overwhelm such limitations, in this paper, we investigated the functional connectivity of brain regions by addressing their hierarchical integration at multiple scales (from the single region level to clusters and entire subnetworks level) within the whole functional network[39] in a cohort of patients diagnosed with SCZ compared to healthy controls (HC). Based on a varied multilevel pipeline, the approach we propose has significant advantages over previously used methods of data-driven brain functional connectivity investigation, in terms of sensitivity of small changes due to small synchronization shifts. We employed recently developed [39] advanced methods of network neuroscience[40,41] to investigate the cerebral region-to-region interaction both in SCZ and HC and the possible dysfunctional reshuffling of connections across the whole network. Briefly, to test the network resilience to disconnection, we ran a percolation analysis by following the disconnection process from the progressive removal of links from the network and the maximum spanning tree (MST) filtration by keeping the most robust links discarding all the others. Finally, we applied an allometric scaling to the MST to quantify its divergence from a linear tree-like organization.

We hypothesize that the altered topology of functional brain networks in SCZ is a consequence of an unbalanced distribution of large and small connectivity strength values. This abnormal connectivity organization includes areas belonging to systems highly engaged in the same processing roles (i.e., areas functionally segregated) and others with different processing assignments (i.e., areas globally integrated)[28]. A possible mechanism underpinning SCZ may arise from a not optimized topological arrangement of the functional network due to the altered connectivity distribution.

We further hypothesize that the cerebral region-to-region interaction is more resistant to disconnection in patients than in healthy subjects due to a dysfunctional reshuffling of significant connections across the whole network. Consequently, the loss of higher cortical hierarchies, responsible for predicting representations, should be a prodromal condition for the reduced specialization of the patients' brain functional connectivity patterns.

## Methods

*Participants*

Forty-four patients diagnosed with SCZ according to the DSM-5[42] were recruited. The clinicians who had been treating the patients and knew their clinical history made the preliminary diagnosis. A senior research psychiatrist (GS) confirmed all prior diagnoses using the Structured Clinical Interview for DSM-5-research version (SCID-5 for DSM-5, Research Version; SCID-5-RV)[43]. Other inclusion criteria were: 1) age between 18 and 65 years; 2) at least eight years of education; 3) no dementia or cognitive deterioration according to the DSM-5 criteria, and a Mini-Mental State Examination score[44] higher than 24 and 4) suitability for a Magnetic Resonance Imaging (MRI) scan. Exclusion criteria were: 1) a history of alcohol or drug dependence or abuse in the last two years; 2) traumatic head injury, 3) any past or present major medical or neurological illness, 4) any other psychiatric disorder or mental retardation diagnosis and 5) MRI evidence of focal parenchymal abnormalities or cerebrovascular diseases. All patients were in a stable clinical compensation phase and were receiving stable oral doses of one or more atypical antipsychotic drugs. Forty HC were also recruited. They did not differ with patients in terms of age, education, and gender. HC were screened for a current or lifetime history of DSM-5 psychiatric and personality disorders using the SCID-5-RV[43] and SCID-5-PD[45], were right-handed, and with normal or corrected-to-normal vision. Table 1 shows HC and SCZ patients' principal demographic characteristics and the associated statistics (two-sample t-test between them).

Participants gave written informed consent to partake in the study after the procedures had been fully explained. The Ethics Committee of the IRCCS Santa Lucia Foundation approved the study protocol which was conducted in accordance with the Helsinki Declaration[46].

**Table1** Principal demographic characteristics of the studied subjects

| | ***Healthy Subjects (HC) (n = 40)*** | ***Schizophrenia Patients (SCZ) (n = 44)*** | ***T value (p)*** |
|---|---|---|---|
| *Gender (F/M)* | *20/20* | *22/22* | - |
| *Age (years) (mean (SD))* | *36.3 (9.7)* | *32.1 (13.1)* | *1.6 (0.11)* |
| *Education (years) (meanSD())* | *13.9 (2.9)* | *12.7 (2.9)* | *1.8 (0.08)* |

*Data acquisition and pre-processing*

FMRI data were collected at 3T (Philips Achieva equipped with a thirty-two channel receive-only head coil) using gradient-echo echo-planar imaging for the registration of the blood oxygen level-dependent (BOLD) signal (TR = 3 s, TE = 30 ms, matrix = 80 x 80, FOV=224x224, slice thickness = 3 mm, flip angle = 90Aˆ◦, 50 slices, 240 vol).

We also acquired high-resolution T1-weighted whole-brain structural scans (1x1x1 mm voxels). Subjects were asked to keep their eyes open, and their cardiac and respiratory cycles were recorded using the scanner's built-in photoplethysmograph and a pneumatic chest belt, respectively. For each subject and each time-series, we removed possible sources of physiological variance: time-lock cardiac and respiratory artifacts through linear regression[47] (i.e., two cardiac and respiratory harmonics, respectively, together with four interaction terms). We also looked for the effect of low-frequency respiratory and heart rates[48–50]. The pre-processing of fMRI data consisted of: head-motion and slice timing corrections plus the discard of voxels not belonging to the brain (with FSL: FMRIB's Software Library, www.fmrib.ox.ac.uk/fsl). We used head motion parameters estimation to obtain the Framewise Displacement (FD). The subjects were included in this study only if they presented a Framewise Displacement smaller than 0.5 mm; however, time points with high FD (FD > 0.2 mm) were replaced using a least-squares spectral decomposition following[51].

Then, we detrended, demeaned, and band-pass filtered (frequency range 0.01-0.1 Hz) data using custom Matlab algorithms. We performed a 2-step registration in line with group-analysis: we first transformed fMRI data from the functional space to the structural space with FLIRT (FMRIB's Linear Registration Tool) for each subject. Using Advanced Normalization Tools (ANTs; Penn Image Computing & Science Lab, http://www.picsl.upenn.edu/ANTS/), the data were subsequently non-linearly sent to the standard space (Montreal Neurological Institute MNI152 standard map).

A spatial smooth was applied to the data (5x5x5 mm full-width half-maximum Gaussian kernel).

- Please Insert Figure1 about here –

*Network analysis*

The whole network analysis was characterized by several steps hereby summarized (all the steps are graphically reported as a flowchart in Figure 1): first, we built the brain functional network by computing the correlation matrices among brain areas; then, we calculated the variability of the correlation matrices across subjects in order to evaluate the inter-subject variability; subsequently, we ran a percolation analysis and applied a MST filtration in order to test network resilience. Briefly, percolation analyses can elucidate network redundancy or robustness where, for example, apparently redundant nodes provide alternative dispersal pathways should any nodes be lost or impacted. MST filtration was computed by keeping the most robust links discarding all the others. Finally, we applied an allometric scaling to the MST in order to study its topological properties.

*Network construction*

We used an AAL mask[52] to parcellate the human brain in 58 anatomical regions for each hemisphere. The fMRI signals were extracted from each voxel and averaged in each region of interest (ROI), ending with 116 BOLD time-series. Then, Pearson's correlation coefficient was used to estimate time-series pairwise similarity, which led to a full and symmetric correlation matrix.

The debate on the meaning of negative correlation values in fMRI studies is still open[53] without a consensus supporting their inclusion or exclusion in the analysis. Since they do not bring any relevant information to the actual study's scope, each correlation coefficient was transformed in Z-values[54] and squared, returning a weighted adjacency matrix associated with a weighted fully connected graph.

*Inter-subject Variability*

To evaluate the variability of the correlation matrices across subjects, for each diagnostic group (i.e., HC and SCZ), the variation coefficient was computed as the standard deviation ratio to the mean across subjects of the correlation values between each pair of brain areas. The coefficient of variation distribution across ROIs was fitted to a log-normal function. The Mann-Whitney-Wilcoxon test[55] was used to address their statistical difference, with the number of degrees of freedom equal to the number of links in the network ($N(N-1)/2$, $N = 116$).

*Percolation Analysis*

A subject-wise percolation analysis[39] for the two groups was performed. Given the individual weighted adjacency matrix, its entries were ranked in ascending order. One at a time, the links corresponding to the ordered list values (i.e., the corresponding weights) were removed by deleting them from the rank. The number of connected components and the size of the largest one (Giant Component) were evaluated progressively after each link was removed (i.e., at each percolation threshold). The percolation curves, intended as the number of connected components as a function of the weight removed, were averaged across subjects. Results are reported together with the 95% confidence interval. The size of the giant component across subjects and the node degree variation across ROIs were investigated to quantify the difference between the two percolation processes. We estimated the lines of reduction as the boundaries over which the giant component's size resulted reduced by 5% to 50% of its original size (namely 116). For each line of reduction, the area under the curve (AUC) was calculated. The difference between the areas corresponding to two consecutive curves, ΔAUC, normalized to the AUC associated with the maximum giant component size ($AUC_{MaxGCSize}$), was computed in the two groups. The rate of change of the two trends was established by fitting them to the exponential decay. As far as the comparison between HC and SCZ ROIs degree at different percolation thresholds, we calculated the square root of the sum of the squared differences between consecutive degree values across the thresholds (Degree Variation Coefficient):

$\Delta d^R = [\sum_{t=1}^{N}(d_{t+1}^R - d_t^R)^2]^{\frac{1}{2}}$, where t is the threshold index, and R is the ROI index. The distribution density of these values has been fitted to a log-normal curve. The statistical difference between the two distributions (HC and SCZ) was established through the Mann-Whitney-Wilcoxon test[55], with the number of degrees of freedom equal to the number of links in each network (N(N-1)/2, N = 116).

*Maximum Spanning Tree*

We computed the Maximum Spanning Tree (MST) for each brain network, keeping each node's maximum weighted link and discarding all the others. The resulting network components were connected through only one connection: the strongest one not forming cycles[39].

- Please Insert Figure2 about here –

*Allometric Scale*

Once the MST was computed for each node, two quantities were estimated: (i) $A_i$, the number of nodes forming the subtree having node i as root (including i); and (ii) $C_i = \sum_k A_k$, where k runs over all nodes in the subtree having root i (including i). It has been observed that in many cases, there exists a power-law relation between these two quantities, $C \propto A^{\eta}$, with the exponent universally identified for food-webs, river, and vascular networks[56,57]. In general, the exponent η, which characterizes the allometric relation of a planar tree, ranges between 1 and 2, where $\eta \rightarrow 1$ for a star-like topology and $\eta = 2$ for a one-dimensional chain-like structure. Thus, the eventual observation of $\eta$ close to 1 (2) is the signature of a global star-like (chain-like) structure of the whole tree. The MSTs associated with the human (correlation) functional brain networks are undirected by construction[39,58].

Thus, to compute the quantities mentioned above ($A_i$, $C_i$) for each node i, we have artificially introduced a directionality by choosing one ROI of the undirected tree as the root, determining the different paths towards the remaining ROIs. In Figure2(a)-(d), we show three examples of such directionality on a toy model composed of ten nodes related to three possible choices of the root: nodes number 5, 7, and 1. Furthermore, in Figure2(e), we give a pictorial representation of the directed version of the MST of the human functional brain network once the choice of a particular node (ROI 51) as root was made. We considered all the possible directed versions, considering each node as root and associating to the MST the induced direction. In other words, we built 116 replicas of the MSTs related to each subject. Such MSTs have the same list of edges and weights, with different links direction.

- Please Insert Figure3 about here –

**Results**

*Inter-subject variability*

Figure1 shows the coefficient of variation for HC and SCZ. According to the analysis of the two cohorts of subjects, HC (Figure3(a)) returned a level of homogeneity much larger than the one observed for SCZ (Figure3(b)). Specifically, the distribution of the coefficients of variation across ROIs was fitted to a log-normal function both for HC and SCZ, with the following mean values and standard deviations with the relative errors of curve fitting: [($\mu_{HTH} \pm \Delta\mu_{HTH}$) = (0.74±0.01), ($\sigma_{HTH} \pm \Delta\sigma_{HTH}$) = (0.21±0.01)] and [($\mu_{SCZ} \pm \Delta\mu_{SCZ}$) = (1.41±0.01), ($\sigma_{SCZ} \pm \Delta\sigma_{SCZ}$) = (0.85±0.01)] (Figure3(c)). According to a Mann-Whitney-Wilcoxon test, the two distributions were found significantly different ($Z = 57$, $p<.001$).

Since this outcome makes meaningless the analysis of an "average brain" for HC and SCZ, we performed a subject-wise study, showing the average and the 95% confidence interval of the analysed quantities .

- Please Insert Figure4 about here –

*Percolation Analysis*

For each subject separately, we evaluated the percolation process in the functional network[39,58]. Figure4 shows the average percolation curve computed for the two groups (HC and SCZ) together with the 95% confidence interval. A net separation emerges between the average percolation curves of the two groups. Specifically, the functional brain network in SCZ appears more resistant to a disgregation process induced by the removal of weak links. Accordingly, the number of connected components remains smaller than the number observed in HC in the range [0.2, 0.8] of thresholds. It suggests a stronger resistance to the decomposition in separate components of the global network architecture for the SCZ group.

- Please Insert Figure5 about here –

The size of the giant component within each network (HC and SCZ) was estimated across thresholds (Figure5(a)-(b)) to estimate the source of such difference in the percolation processes quantitatively. Firstly, the area under the curve associated with the maximum giant component size ($AUC_{MaxGCSize}$) normalized to the total area shows that the SCZ group is more resistant to the initial fragmentation than the HC one ($AUC_{SCZ}$ /$AUC_{total}$ = 0.30 and $AUC_{HC}$ /$AUC_{total}$ = 0.16). It implies a delay at the beginning of the disgregation process between SCZ and HC. Moreover, the estimation of the disconnection rate (Figure5(c)) demonstrates that, even if the process starts later in SCZ, its progression is faster, pairing the two groups when the giant component size is halved. The difference between the areas under two consecutive curves, ΔAUC, normalized to the area under the curve associated with the maximum giant component size, $AUC_{MaxGCSize}$, at different values of the giant component size reduction was fitted to the exponential decay. The speed of decay (SoD) was found to be $SoD_{HC} = (5.03 \pm 0.02)$ and $SoD_{SCZ} = (3.11 \pm 0.01)$.

- Please Insert Figure6 about here –

We also checked if the differences between the two percolation processes can be ascribed to a different distribution of the weights within each network. Figure6 shows the correlation values of the networks of all subjects pooled together (squared values in the inset). Even if the distribution of the weights collected from SCZ is wider than the one coming from HC, not significant differences were found according to a Mann-Whitney-Wilcoxon test. On the contrary, by comparing how each ROI degree (averaged across subjects) changes at each threshold value (Figure7(a-b)), we discovered relevant differences between the two groups. Specifically, the distributions of the Degree Variation Coefficient for HC and SCZ (Figure7(c)) have been found significantly different ($Z = 51$, $p < .001$) according to a Mann-Whitney-Wilcoxon test.

- Please Insert Figure7 about here –

Accordingly, in the case of SCZ, the node degree decrease appears almost homogeneous for all ROIs. In other terms, our results reveal that during the percolation, weak links are removed almost 'randomly' from the functional brain network of schizophrenic patients, with the consequence of affecting almost in the same way the variation of the degree of each ROI. Simultaneously, for the HC group, nodes disconnect rapidly from the rest of the network and become soon isolated. It suggests that weaker links are uniformly distributed in patients' functional brain networks instead of being concentrated around the same node or groups of nodes as for the healthy subjects.

*Maximum Spanning Tree*

In order to investigate more on the differential functional organization in HC and SCZ, we filtered the correlation network of each individual by extracting its Maximum Spanning Tree (MST)[39]. Both HC and SCZ showed a chain-like arrangement of their MSTs.

The divergence of the MSTs structure from a linear organization was quantified according to the analysis of the allometric exponent. Once the MST from each individual was computed, and the root induced the directions along the tree, the exponent $\eta$ of the allometric scaling law was computed for each individual. This procedure was separately repeated for 116 trees (one for each ROI chosen as root) for both HC and SCZ subjects. Figure8 shows the $\eta$ values for each ROI, averaged across subjects, reported with the 95% confidence interval. The mean allometric exponents for the two groups were the same across roots (Figure8), revealing on average a similar organization of the functional brain network backbone for HC and SCZ.

- Please Insert Figure8 about here –

To investigate further the possible origin of the differences showed by the percolation analysis, we extended the allometric analysis beyond the correlation network backbone. Specifically, we defined higher rank MSTs as follows: first, we rename the correlation network's MST as the *first rank* MST. Then, we identify a *second rank* MST as the one obtained from the original correlation network by eliminating the links belonging to the first rank MST. Accordingly, we can define the nth rank MST as the MST of the correlation network from which we have already removed all the previous n-1 MSTs. We computed the first four MSTs for each ROI used as root and for each of the subjects. Figure9 shows that, starting from a similar condition, allometric exponents tend to reduce as the MST rank increases. However, the different speed of reduction (SR) determines an evident and growing difference between the two groups of subjects, with the allometric exponents decreasing more slowly for SCZ than HC. In fact, an exponential fit ($\eta(Rank) = (\eta_0 - \eta_\infty)*exp(-Rank/SR)+ \eta_\infty$) of the average η across the ROIs returned $SR_{HC} = (0.610 \pm 0.001)$ and $SR_{SCZ} = (0.802 \pm 0.001)$, for HC and SCZ respectively.

- Please Insert Figure9 about here –

**Discussion**

Here we report an advanced network-based analysis of fMRI data recorded at rest in SCZ and HC. We aimed to assess possible alterations in the brain's basal functional network organization of SCZ through percolation and MST analyses [39,58]. Our findings demonstrate a global change of the connectivity strength distribution in the functional networks of SCZ as compared to HC, consisting of an increased homogeneity of the distribution of the weighted links across the whole network. Specifically, from the inspection and analysis of the percolation curves, we found that the functional networks in SCZ are characterized by a region-to-region interaction which is more resistant to disconnection than in HC. Indeed, by scrolling the threshold (below which links are erased) from low to high connectivity strength values, the number of separate clusters is systematically smaller in SCZ. Similarly, we found that, when comparing the size of the giant component[59] as a function of the threshold, the size reduction in HC is always faster than that observed in SCZ. A wider distribution of the connectivity strengths in SCZ cannot explain such differences in the percolated networks, mainly because the two distributions become very similar when the squared correlation values are considered.

On the other hand, analysing the distribution of the node degree computed for each percolated network, a clear discrepancy was found between the two groups with consequent delayed emergence of a second connected component in the percolating network. This outcome sheds light on the existence of a more homogeneous distribution of connections in the functional architecture of brains of patients diagnosed with SCZ compared to HC. Since the modular structure of a network is determined by the unbalance or predominance of inward and outward links of individual communities, rather than by the average total distribution of edge weights, an effect of the re-arrangement of the weights may result in a progressive loss of hierarchy of functional connections, which, in turns, leads to an alteration or disruption of modularity.

Modularity structure modifications in SCZ were reported as due to changes in community participation of nodes in the somatosensory, subcortical, auditory, default mode, and salience networks[35] and as a specific fragmentation of somatosensory cortices[36]. Nonetheless, the local specificity of the aberrant cerebral functioning in SCZ is still under debate. Hypo-connectivity was reported in the olfactory cortex, temporal pole, angular gyrus, parahippocampus, amygdala, caudate, and pallidum[28,32]. Conversely, hyper-connectivity was found within the default mode network[60], the bilateral striatum[61], and between the default mode and visual and motor regions[62]. These findings suggest a widespread modulation of connectivity in SCZ that involves both directions (i.e., increases and decreases) and are in line with our results: functional brain networks in SCZ are characterized by a more homogeneous distribution of weights, where strong correlation patterns and weak ones share the same topology. Analogous conclusions were reached by authors[28], who pointed to the altered graph topology in SCZ as responsible for changes in the brain's activity and connectivity complexity.

Conversely, the MST analysis of the functional network does not differ between the two groups, thus implying that the backbone of the functional organization seems to be very similar between groups [39]. A quantitative confirmation of this result is given by the allometric exponent calculated for each subject. MSTs derived from functional brain networks can, in principle, range between two different topologies: the linear configuration (allometric exponent equal to 2), the least efficient with the lowest cost of functioning, and the star-like organization (allometric exponent that tends to 1), the most efficient but with the highest cost of functioning[63]. The configuration of the ones we derived from functional brain networks is associated with an allometric exponent approximately equal to 1.5, a configuration that balances efficiency and cost. In the context of functional brain connectivity, which involves interaction between regions not necessarily close one to the other, and where the proximity is just a consequence of their synchronization, efficiency and cost are intended as the ability to organize information transfer via a synchronous coordination[64] and the provision of energetic resources that sustain such organization[65] respectively. Since we observed a widespread change in the distribution of connectivity strength in brain networks of SCZ along with an almost unaltered MST topology, we explored the allometric properties of the MSTs at increasing rank. Our findings show that by progressively removing MSTs from a functional network, the value of the allometric exponent decreases both in SCZ and in HC; however, the rate of reduction of the exponent is faster in HC, leading to a net separation of the two groups at the third rank MST. It suggests that not only the hierarchy and modular structure of the functional network are reduced by a homogeneous distribution of the connectivity strengths but also that in terms of the global integration of brain regions in the network, weaker links in SCZ guarantee the same topology shown by the stronger ones [28]. The higher topological similarity of the MSTs of different ranks, observed in SCZ as compared to HC, suggests that while a given stimulus can engage a single functional connectivity path[39] in patients, it determines the simultaneous involvement of different ones. A possible explanation of this conclusion can be found in the disconnection hypothesis[8,66], which relates the molecular and the neuronal pathophysiology to SCZ symptoms. It states that this psychiatric disorder may stem from an abnormal response of the NMDA receptor to specific neuromodulatory receptor activation. A failure of such mechanism may lead to an inability to modulate the congruity of sensory evidence, corresponding to the congruity of beliefs about the causes of sensory cues, and consequently to false inference (e.g., hallucinations and delusions)[66]. Accordingly, our findings show that a reshuffling of connectivity strengths with the realization of alternative connectivity patterns occurs in the illness that disrupts the local modular organization (i.e., the somatosensory community). A consequence of this multi-choice configuration is the loss of higher levels of cortical hierarchies that generate predictions of representations in lower levels, jeopardizing in SCZ the brain's ability to process sensory information by optimizing explanations for its sensations[67–69]. It is worth further commenting on the possible physiological causes of the heterogeneity reduction, and the loss of hierarchy showed by the correlation strength observed in SCZ. In a computational model[70], it was proposed that the best way to obtain functional networks with topological properties matching those reported experimentally in SCZ would be the decrease of the strength of excitatory synaptic inputs as a consequence of the disruption of synaptic mechanisms. At the neurophysiological level, the disconnection hypothesis accounts for such altered mechanisms[9]. Nonetheless, an excess of synaptic refinement (enhanced pruning) has also been hypothesized to underlie the neuropathology of SCZ[71,72]. Our results suggest that the brain activity in SCZ is characterized by a subtle change of the global functional architecture, which is not random but involves both an increase and a decrease of the local connectivity strengths. It probably occurs due to the brain's attempt to compensate for an imbalance of the local homeostatic signaling [73]. Such untargeted compensation may homogenize the patterns of information spread in the brain, which is expected to reach all areas without the whole system's involvement in the activity[74].

In summary, this study indicates how previously reported fragmentation of the modular structure of functional connectivity in medicated patients diagnosed with SCZ might be due to a redistribution, and consequently a homogenization, of the connectivity strengths among all the regions of the brain. Our findings support the theory that aberrant connectivity may induce deficits that propagate to higher functions through a bottom-up process[66]. Moreover, we report several equivalent basal functional skeletons in SCZ (MSTs), which implies the lack of specificity in the brain's functional connectivity organization that is actually expected to be hierarchically set up to account for the correct integration for the proper functioning.

**Captions**

Figure1: *Flowchart of the network analysis*. The temporal progression of the network analysis can be divided into three main parts: the network construction from BOLD time-series averaged in the regions of interest of the AAL atlas, the percolation analysis that highlights the integration process of the network's nodes into modules according to the strength of their connectivity and finally the unpacking of the whole brain graph in terms of Maximum Spanning Trees (MSTs) that allows one to identify deviations from the hierarchical organization of the brain.

Figure2: *Allometric scale*. (a) A toy model of 10 nodes. According to the different nodes chosen as roots, it is possible to obtain ten directed versions of the undirected tree (a). (b)-(d) three directed versions of the tree in (a) with roots equal to, respectively, nodes 5, 7, and 1. Numbers in brackets represent the quantities ($A_i$, $C_i$) associated with each node. (e) Maximum spanning tree of the human functional brain network of an individual chosen at random in healthy subjects with directionality induced by choosing node fifty-one as root. Colors in (e) represent anatomical regions according to the grouping of AAL parcellation shown on the right bottom: • Frontal Lobe; • Insula; • Cingulate; • Temporal Lobe; • Occipital Lobe; • Parietal Lobe; • Deep Grey Matter; • Cerebellum.

Figure3: *Inter-subjects variability*. (a) Coefficients of variation matrix for HC, (b) Coefficients of variation matrix for SCZ, and (c) density of the distribution of the pairwise coefficients of variation for HTH (blue) and SCZ (red). The coefficient of variation was calculated as the standard deviation ratio to the mean across subjects of the correlation values between each pair of brain areas. Data were fitted to a log-normal distribution with mean μ and variance σ2. The curve fitting results are reported for both HC and SCZ with the errors of the fitting procedure. a
According to a Mann-Whitney-Wilcoxon test, the two distributions resulted significantly different ($Z = 57$, $p < .001$).

Figure4: *Percolation Analysis*. The number of connected components of the percolated network versus the related correlation threshold for HC (blue) and SCZ (red). The two curves represent the average of the individual percolation curves and are reported with a 95% confidence interval.

Figure5: *Percolation and giant component analysis*. (a) Giant component size variation for HC, (b) Giant component size variation for SCZ, and (c) The difference between the areas under two consecutive curves, ΔAUC, normalized to the area under the curve associated with the maximum giant component size, $AUC_{MaxGCSize}$ (yellow line), for HC (blue) and SCZ (red). $\Delta AUC/AUC_{MaxGCSize}$, for different values of the giant component size reduction, was fitted to the exponential decay. The speed of decay (SoD) was found to be $SoD_{HC}$ = (5.03 ± 0.02) and $SoD_{SCZ}$ = (3.11 ± 0.01) for HC and SCZ, respectively. Dotted lines represent 95% of prediction bounds.

Figure6: *Weights distribution comparison*. The density of the distributions of the human functional brain networks' correlation values for healthy subjects (blue) and schizophrenic patients (red). Inset: distribution of the squared correlation values.

Figure7: *Percolation and node degree analysis*. (a) Variation of the node degree averaged across subjects for HC, (b) Variation of the node degree averaged across subjects for SCZ (c) density of the distribution of the Degree Variation Coefficients for HC (blue) and SCZ (red). Data were fitted to a log-normal distribution with mean μ and variance $\sigma^2$. The curve fitting results are reported for both HC and SCZ with the errors of the fitting procedure. According to a Mann-Whitney-Wilcoxon test, the two distributions resulted significantly different ($Z = 51$, $p < .001$).

Figure8: *Allometric exponent for the first rank MST*. Mean allometric exponent $\eta$, separately averaged over the two groups HC and SCZ, with 95% confidence interval for the first rank MST. Each ROI was chosen as a directionality root.

Figure9: *Allometric exponent for higher-order MSTs*. Mean allometric exponent $\eta$, separately averaged over the two groups HC and SCZ, with 95% confidence interval for the second, third and fourth rank MSTs. The speed of reduction (SR) of the allometric exponent is larger for SCZ than for HC. An exponential fit ($\eta(rank) = (\eta_0 - \eta_\infty)*\exp(-rank/SR)+ \eta_\infty$) of the average $\eta$ across the ROIs returned $SR_{HC} = (0.61\pm0.08)$ $SR_{SCZ} = (0.80\pm0.10)$. Each ROI was chosen as a directionality root.

**Authors contribution**

RM performed the analysis, prepared the figures, and wrote the manuscript. FP and NB tested patients, acquired the data, and wrote the manuscript. AG designed the study, discussed and interpreted the results, and wrote the manuscript. GC discussed and interpreted results. GS designed the study, recruited and assessed patients, supervised the paper. TG designed the study, acquired and processed the data, interpreted the results, prepared the figures, and wrote the manuscript. All authors edited the text and approved the final version.

## References

1. Fox, M. D., Snyder, A. Z., Vincent, J. L. & Raichle, M. E. Intrinsic fluctuations within cortical systems account for intertrial variability in human behavior. *Neuron* **56**, 171–184 (2007).
2. Biswal, B. B. *et al.* Toward discovery science of human brain function. *Proc. Natl. Acad. Sci.* **107**, 4734–4739 (2010).
3. Venkataraman, A., Whitford, T. J., Westin, C.-F., Golland, P. & Kubicki, M. Whole brain resting state functional connectivity abnormalities in schizophrenia. *Schizophr. Res.* **139**, 7–12 (2012).
4. Rashid, B. *et al.* Classification of schizophrenia and bipolar patients using static and dynamic resting-state fMRI brain connectivity. *Neuroimage* **134**, 645–657 (2016).
5. Sheffield, J. M. & Barch, D. M. Cognition and resting-state functional connectivity in schizophrenia. *Neurosci. Biobehav. Rev.* **61**, 108–120 (2016).
6. Lewandowski, K. E. *et al.* Functional connectivity in distinct cognitive subtypes in psychosis. *Schizophr. Res.* (2018).
7. Weinberger, D. R., Berman, K. F. & Illowsky, B. P. Physiological dysfunction of dorsolateral prefrontal cortex in schizophrenia: III. A new cohort and evidence for a monoaminergic mechanism. *Arch. Gen. Psychiatry* **45**, 609–615 (1988).
8. Spalletta, G. *et al.* Chronic schizophrenia as a brain misconnection syndrom: a white matter voxel-based morphometry study. *Schizophr. Res.* **64**, 15–23 (2003).
9. Stephan, K. E., Friston, K. J. & Frith, C. D. Dysconnection in schizophrenia: from abnormal synaptic plasticity to failures of self-monitoring. *Schizophr. Bull.* **35**, 509–527 (2009).
10. Singer, W. Neuronal synchrony: a versatile code for the definition of relations? *Neuron* **24**, 49–65 (1999).

11. Varela, F., Lachaux, J.-P., Rodriguez, E. & Martinerie, J. The brainweb: phase synchronization and large-scale integration. *Nat. Rev. Neurosci.* **2**, 229 (2001).

12. Gregoriou, G. G., Gotts, S. J., Zhou, H. & Desimone, R. Long-range neural coupling through synchronization with attention. *Prog. Brain Res.* **176**, 35–45 (2009).

13. Singer, W. Distributed processing and temporal codes in neuronal networks. *Cogn. Neurodyn.* **3**, 189–196 (2009).

14. Fries, P. Rhythms for cognition: communication through coherence. *Neuron* **88**, 220–235 (2015).

15. Gili, T., Ciullo, V. & Spalletta, G. Metastable states of multiscale brain networks are keys to crack the timing problem. *Front. Comput. Neurosci.* **12**, 1–8 (2018).

16. Boksa, P. Abnormal synaptic pruning in schizophrenia: Urban myth or reality? *J. psychiatry Neurosci. JPN* **37**, 75 (2012).

17. Stoneham, E. T., Sanders, E. M., Sanyal, M. & Dumas, T. C. Rules of engagement: factors that regulate activity-dependent synaptic plasticity during neural network development. *Biol. Bull.* **219**, 81–99 (2010).

18. Paolicelli, R. C. *et al.* Synaptic pruning by microglia is necessary for normal brain development. *Science (80-. ).* 1202529 (2011).

19. Faludi, G. & Mirnics, K. Synaptic changes in the brain of subjects with schizophrenia. *Int. J. Dev. Neurosci.* **29**, 305–309 (2011).

20. Meyer-Lindenberg, A. *et al.* Evidence for abnormal cortical functional connectivity during working memory in schizophrenia. *Am. J. Psychiatry* **158**, 1809–1817 (2001).

21. Meyer-Lindenberg, A. S. *et al.* Regionally specific disturbance of dorsolateral prefrontal--hippocampal functional connectivity in schizophrenia. *Arch. Gen. Psychiatry* **62**, 379–386 (2005).

22. Esslinger, C. *et al.* Neural mechanisms of a genome-wide supported psychosis variant. *Science (80-. ).* **324**, 605 (2009).

23. Lynall, M.-E. *et al.* Functional connectivity and brain networks in schizophrenia. *J. Neurosci.* **30**, 9477–9487 (2010).

24. Skudlarski, P. *et al.* Brain connectivity is not only lower but different in schizophrenia: a combined anatomical and functional approach. *Biol. Psychiatry* **68**, 61–69 (2010).

25. Fitzsimmons, J., Kubicki, M. & Shenton, M. E. Review of functional and anatomical brain connectivity findings in schizophrenia. *Curr. Opin. Psychiatry* **26**, 172–187 (2013).

26. Li, P. *et al.* Altered Brain Network Connectivity as a Potential Endophenotype of Schizophrenia. *Sci. Rep.* **7**, (2017).

27. Ding, Y. *et al.* Enhanced global-brain functional connectivity in the left superior frontal gyrus as a possible endophenotype for schizophrenia. *Front. Neurosci.* **13**, (2019).

28. Bassett, D. S., Nelson, B. G., Mueller, B. A., Camchong, J. & Lim, K. O. Altered resting state complexity in schizophrenia. *Neuroimage* **59**, 2196–2207 (2012).

29. Uhlhaas, P. J. Dysconnectivity, large-scale networks and neuronal dynamics in schizophrenia. *Curr. Opin. Neurobiol.* **23**, 283–290 (2013).

30. Uhlhaas, P. J. & Singer, W. Abnormal neural oscillations and synchrony in schizophrenia. *Nat. Rev. Neurosci.* **11**, 100 (2010).

31. van den Heuvel, M. P. *et al.* Abnormal rich club organization and functional brain dynamics in schizophrenia. *JAMA psychiatry* **70**, 783–792 (2013).

32. Nelson, B. G., Bassett, D. S., Camchong, J., Bullmore, E. T. & Lim, K. O. Comparison of large-scale human brain functional and anatomical networks in schizophrenia. *NeuroImage Clin.* **15**, 439–448 (2017).

33. Damaraju, E. *et al.* Dynamic functional connectivity analysis reveals transient states of dysconnectivity in schizophrenia. *NeuroImage Clin.* **5**, (2014).

34. Ma, S., Calhoun, V. D., Phlypo, R. & Adali, T. Dynamic changes of spatial functional network connectivity in healthy individuals and schizophrenia patients using independent vector analysis. *Neuroimage* **90**, (2014).

35. Lerman-Sinkoff, D. B. & Barch, D. M. Network community structure alterations in adult schizophrenia: identification and localization of alterations. *NeuroImage Clin.* **10**, 96–106 (2016).

36. Bordier, C., Nicolini, C., Forcellini, G. & Bifone, A. Disrupted modular organization of primary sensory brain areas in schizophrenia. *NeuroImage Clin.* **18**, 682–693 (2018).

37. Ray, K. L. *et al.* Functional network changes and cognitive control in schizophrenia. *NeuroImage Clin.* **15**, (2017).

38. He, H. *et al.* Reduction in gray matter of cerebellum in schizophrenia and its influence on static and dynamic connectivity. *Hum. Brain Mapp.* **40**, (2019).

39. Mastrandrea, R. *et al.* Organization and hierarchy of the human functional brain network lead to a chain-like core. *Sci. Rep.* (2017). doi:10.1038/s41598-017-04716-3

40. Bassett, D. S. & Sporns, O. Network neuroscience. *Nature Neuroscience* (2017). doi:10.1038/nn.4502

41. Bullmore, E. T. & Bassett, D. S. Brain Graphs: Graphical Models of the Human Brain Connectome. *Annu. Rev. Clin. Psychol.* **7**, 113–140 (2011).

42. American-Psychiatric-Association. *Diagnostic and statistical manual of mental disorders (DSM-5)*. (American Psychiatric Publishing, 2013).

43. First, M. B., Williams, J. B. W., Karg, R. S. & Spitzer, R. *Structured Clinical Interview for DSM-5 Research Version (SCID-5 for DSM-5, Research Version; SCID-5-RV).* (Arlington, VA: American Psychiatric Association, 2015).

44. Folstein, M. F., Folstein, S. E. & McHugh, P. R. 'Mini-mental state'. A practical method for grading the cognitive state of patients for the clinician. *J. Psychiatr. Res.* (1975). doi:10.1016/0022-3956(75)90026-6

45. First, M. B. *SCID-5-PD: structured clinical interview for DSM-5 personality disorders: includes the self-report screener structured clinical interview for DSM-5 screening personality questionnaire (SCID-5-SPQ)*. (American Psychiatric Association, 2016).

46. World Medical Association declaration of Helsinki: Ethical principles for medical research involving human subjects. *JAMA - Journal of the American Medical Association* **310**, (2013).

47. Glover, G. H., Li, T. Q. & Ress, D. Image-based method for retrospective correction of physiological motion effects in fMRI: RETROICOR. *Magn. Reson. Med.* **44**, 162–167 (2000).

48. Birn, R. M., Diamond, J. B., Smith, M. A. & Bandettini, P. A. Separating respiratory-variation-related fluctuations from neuronal-activity-related fluctuations in fMRI. *Neuroimage* (2006). doi:10.1016/j.neuroimage.2006.02.048

49. Shmueli, K. *et al.* Low-frequency fluctuations in the cardiac rate as a source of variance in the resting-state fMRI BOLD signal. *Neuroimage* (2007). doi:10.1016/j.neuroimage.2007.07.037

50. Chang, C. & Glover, G. H. Effects of model-based physiological noise correction on default mode network anti-correlations and correlations. *Neuroimage* (2009). doi:10.1016/j.neuroimage.2009.05.012

51. Power, J. D. & others. Methods to detect, characterize, and remove motion artifact in resting state fMRI. *Neuroimage* **84**, 320–341 (2014).

52. Tzourio-Mazoyer, N. *et al.* Automated anatomical labeling of activations in SPM using a macroscopic anatomical parcellation of the MNI MRI single-subject brain. *Neuroimage* (2002). doi:10.1006/nimg.2001.0978

53. Duff, E. P., Makin, T., Cottaar, M., Smith, S. M. & Woolrich, M. W. Disambiguating brain functional connectivity. *Neuroimage* **173**, 540–550 (2018).

54. Fisher, R. A. Frequency Distribution of the Values of the Correlation Coefficient in Samples from an Indefinitely Large Population. *Biometrika* (1915). doi:10.2307/2331838

55. Mann, H. B. & Whitney, D. R. On a Test of Whether one of Two Random Variables is Stochastically Larger than the Other. *Ann. Math. Stat.* (1947). doi:10.1214/aoms/1177730491

56. Banavar, J. R., Maritan, A. & Rinaldo, A. Size and form in efficient transportation networks.

*Nature* **399**, 130 (1999).

57. Garlaschelli, D., Caldarelli, G. & Pietronero, L. Universal scaling relations in food webs. *Nature* (2003). doi:10.1038/nature01604

58. Bardella, G., Bifone, A., Gabrielli, A., Gozzi, A. & Squartini, T. Hierarchical organization of functional connectivity in the mouse brain: a complex network approach. *Sci. Rep.* **16**, 32060 (2016).

59. Gallos, L. K., Makse, H. A. & Sigman, M. A small world of weak ties provides optimal global integration of self-similar modules in functional brain networks. *Proc. Natl. Acad. Sci.* **109**, 2825–2830 (2012).

60. Whitfield-Gabrieli, S. *et al.* Hyperactivity and hyperconnectivity of the default network in schizophrenia and in first-degree relatives of persons with schizophrenia. *Proc. Natl. Acad. Sci. U. S. A.* (2009). doi:10.1073/pnas.0809141106

61. Zhuo, C. *et al.* Functional connectivity density alterations in schizophrenia. *Front. Behav. Neurosci.* **8**, 404 (2014).

62. Rashid, B., Damaraju, E., Pearlson, G. D. & Calhoun, V. D. Dynamic connectivity states estimated from resting fMRI Identify differences among Schizophrenia, bipolar disorder, and healthy control subjects. *Front. Hum. Neurosci.* **8**, 897 (2014).

63. Bullmore, E. & Sporns, O. The economy of brain network organization. *Nat. Rev. Neurosci.* **13**, 336–349 (2012).

64. Fries, P. A mechanism for cognitive dynamics: Neuronal communication through neuronal coherence. *Trends Cogn. Sci.* (2005). doi:10.1016/j.tics.2005.08.011

65. Tomasi, D., Wang, G.-J. & Volkow, N. D. Energetic cost of brain functional connectivity. *Proc. Natl. Acad. Sci.* **110**, 13642–13647 (2013).

66. Friston, K. J. The disconnection hypothesis. *Schizophr. Res.* **30**, 115–125 (1998).

67. Friston, K. Hierarchical models in the brain. *PLoS Comput. Biol.* **4**, e1000211 (2008).

68. Bastos, A. M. *et al.* Canonical microcircuits for predictive coding. *Neuron* **76**, 695–711

(2012).

69. Clark, A. Whatever next? Predictive brains, situated agents, and the future of cognitive science. *Behav. Brain Sci.* **36**, 181–204 (2013).

70. Cabral, J. *et al.* Structural connectivity in schizophrenia and its impact on the dynamics of spontaneous functional networks. *Chaos An Interdiscip. J. Nonlinear Sci.* **23**, 46111 (2013).

71. Cole, M. W., Anticevic, A., Repovs, G. & Barch, D. Variable global dysconnectivity and individual differences in schizophrenia. *Biol. Psychiatry* **70**, 43–50 (2011).

72. Tomasi, D. & Volkow, N. D. Mapping small-world properties through development in the human brain: disruption in schizophrenia. *PLoS One* **9**, e96176 (2014).

73. Landek-Salgado, M. A., Faust, T. E. & Sawa, A. Molecular substrates of schizophrenia: homeostatic signaling to connectivity. *Mol. Psychiatry* **21**, 10 (2016).

74. Russo, R., Herrmann, H. J. & de Arcangelis, L. Brain modularity controls the critical behavior of spontaneous activity. *Sci. Rep.* **4**, 4312 (2014).

Figure 1

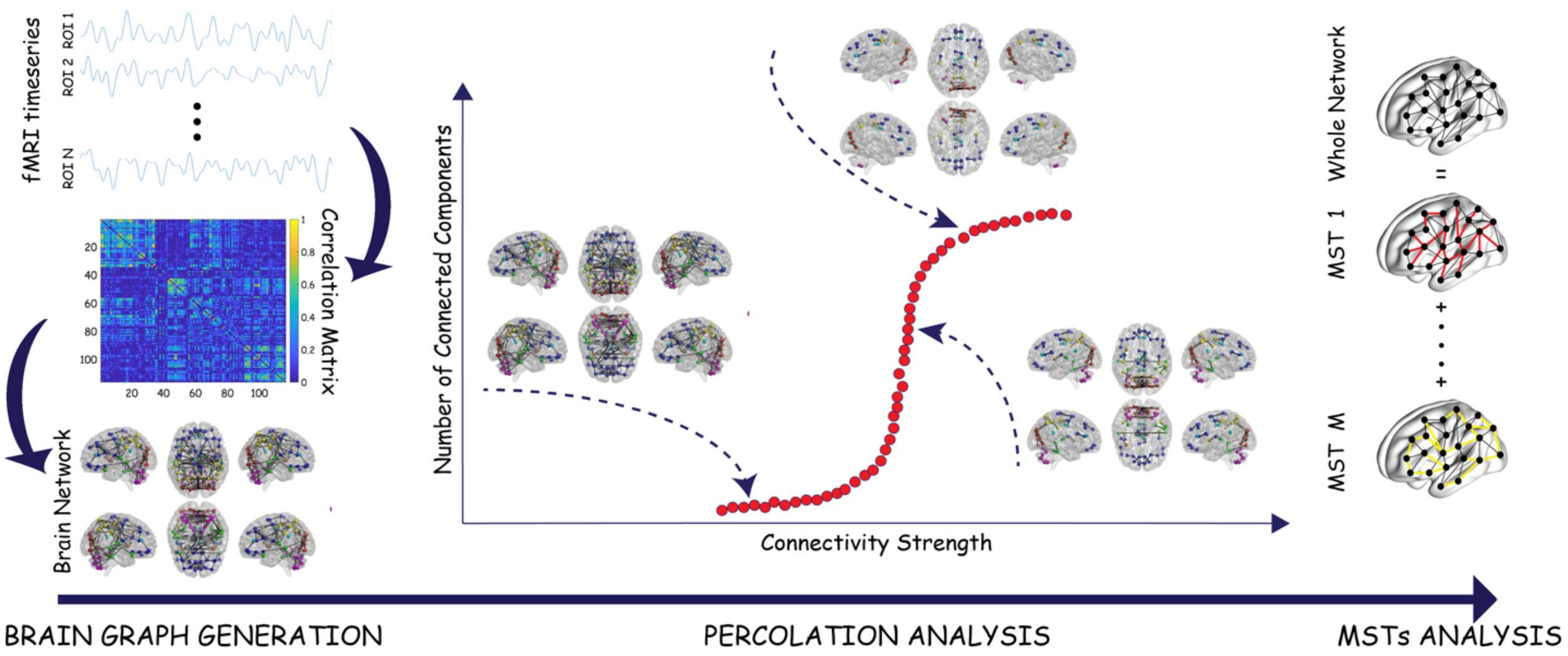
fMRI timeseries
ROI 1
ROI 2
ROI N
Correlation Matrix
Brain Network
Number of Connected Components
Connectivity Strength
Whole Network
MST 1
MST M
BRAIN GRAPH GENERATION
PERCOLATION ANALYSIS
MSTs ANALYSIS

Figure 2

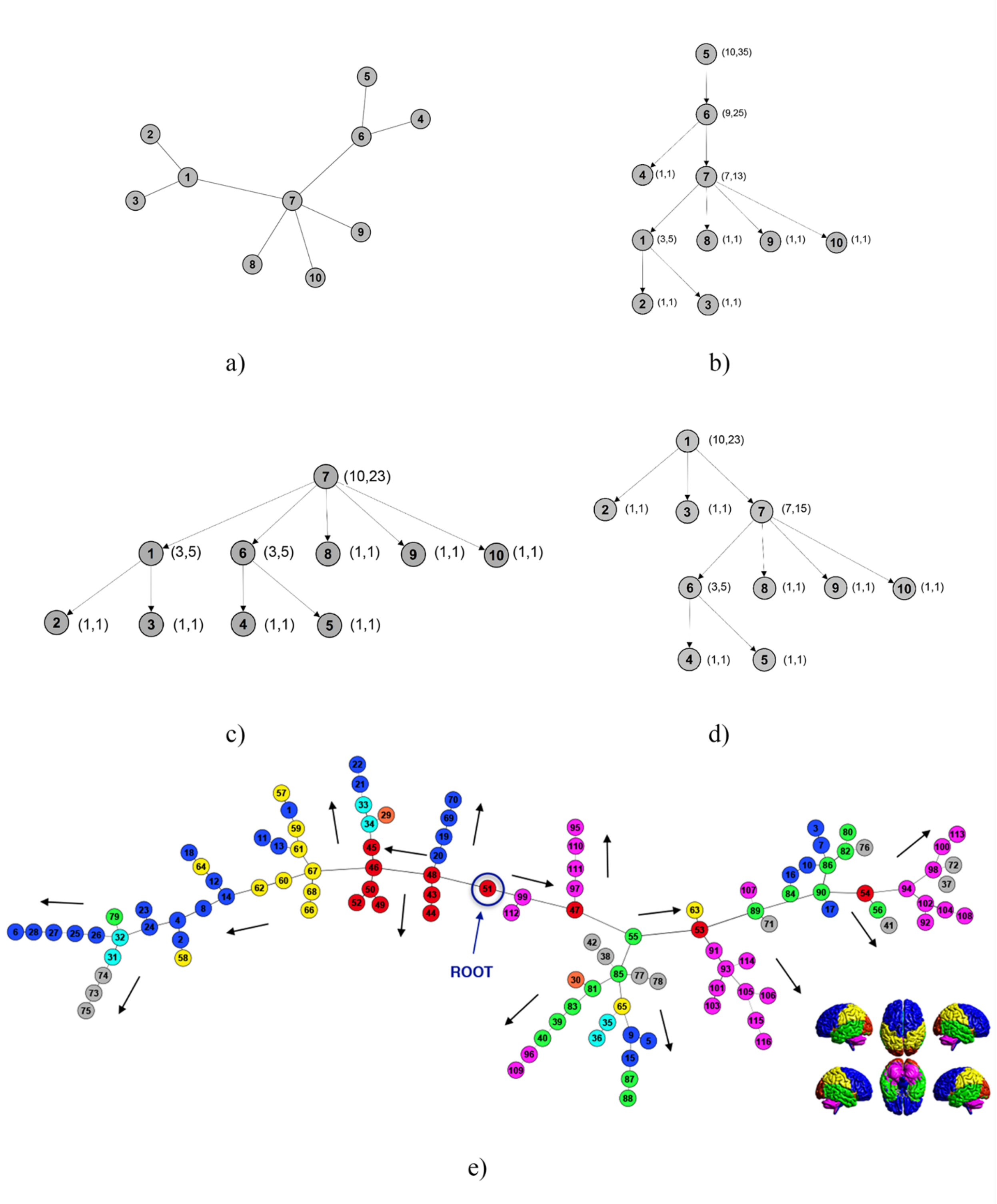

Figure 3

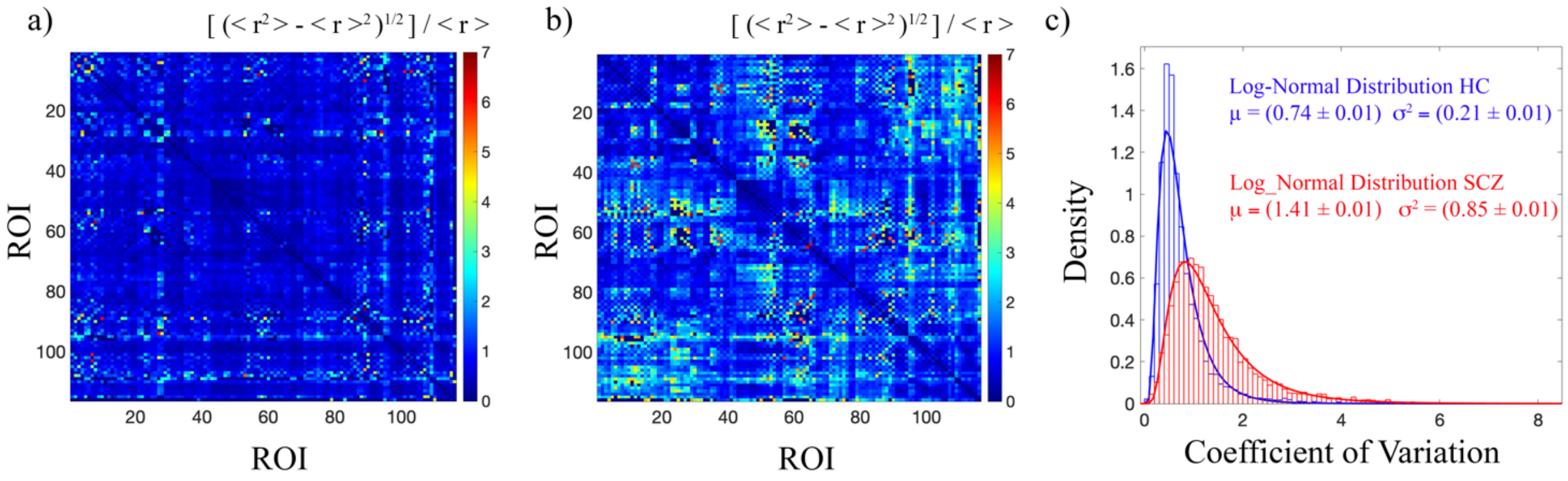
a)
[ (< r² > - < r >²)^1/2 ] / < r >
ROI
ROI
b)
[ (< r² > - < r >²)^1/2 ] / < r >
ROI
ROI
c)
Log-Normal Distribution HC
μ = (0.74 ± 0.01)  σ² = (0.21 ± 0.01)
Log_Normal Distribution SCZ
μ = (1.41 ± 0.01)  σ² = (0.85 ± 0.01)
Density
Coefficient of Variation

Figure 4

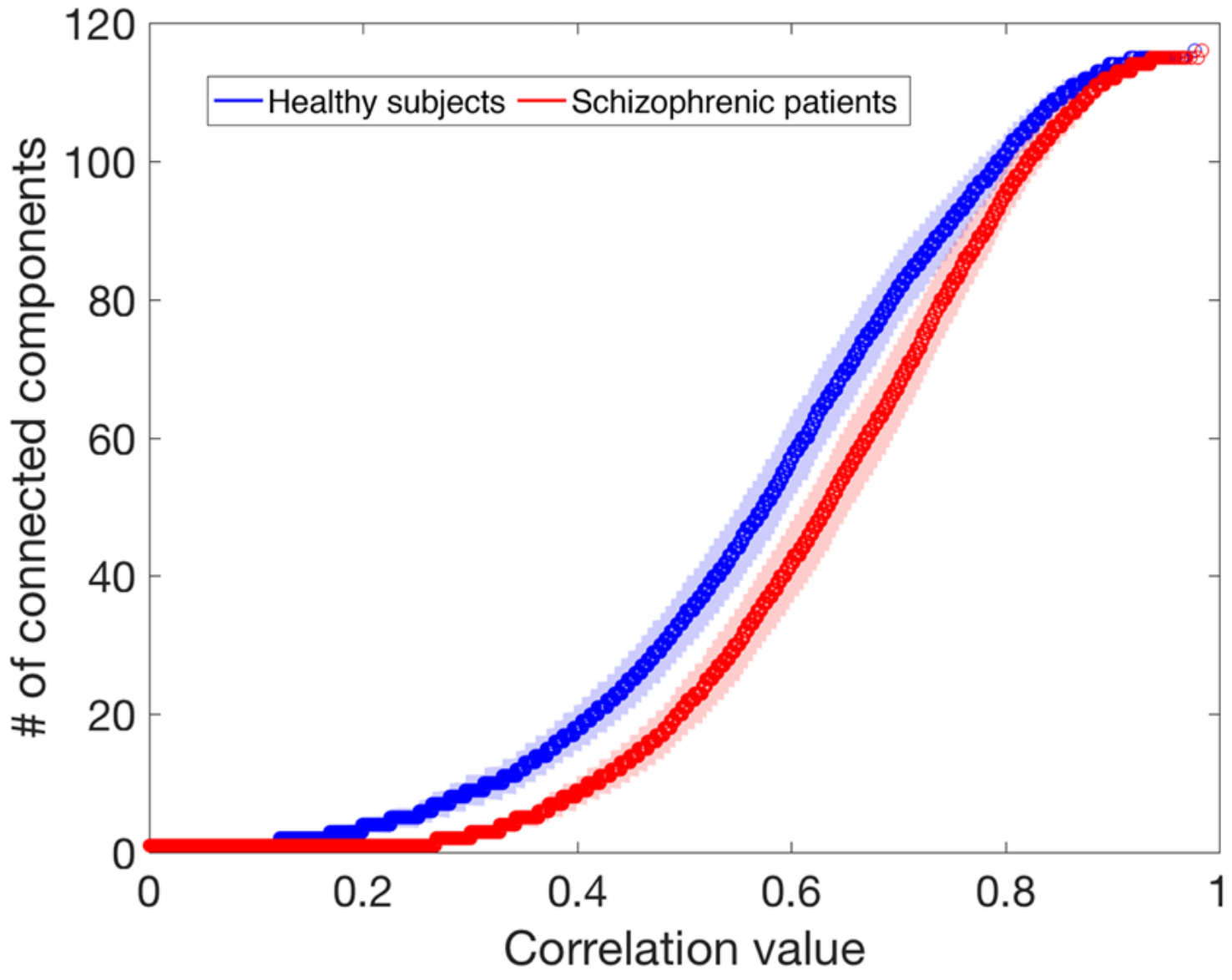
Healthy subjects
Schizophrenic patients
# of connected components
Correlation value
0
20
40
60
80
100
120
0
0.2
0.4
0.6
0.8
1

Figure 5

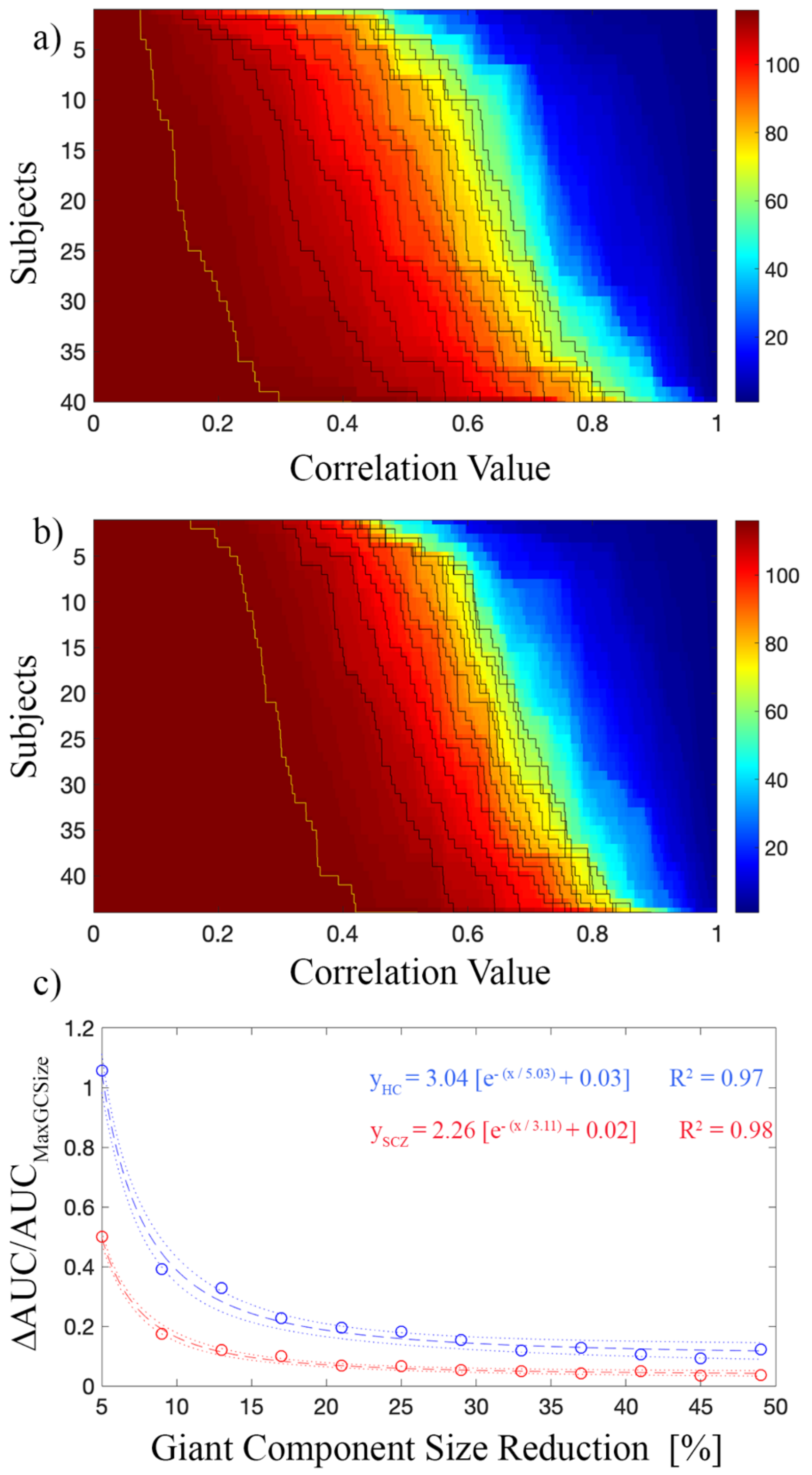
a)
Subjects
Correlation Value
b)
Subjects
Correlation Value
c)
ΔAUC/AUC_MaxGCSize
y_HC = 3.04 [e^-(x / 5.03) + 0.03]
R² = 0.97
y_SCZ = 2.26 [e^-(x / 3.11) + 0.02]
R² = 0.98
Giant Component Size Reduction [%]

Figure 6

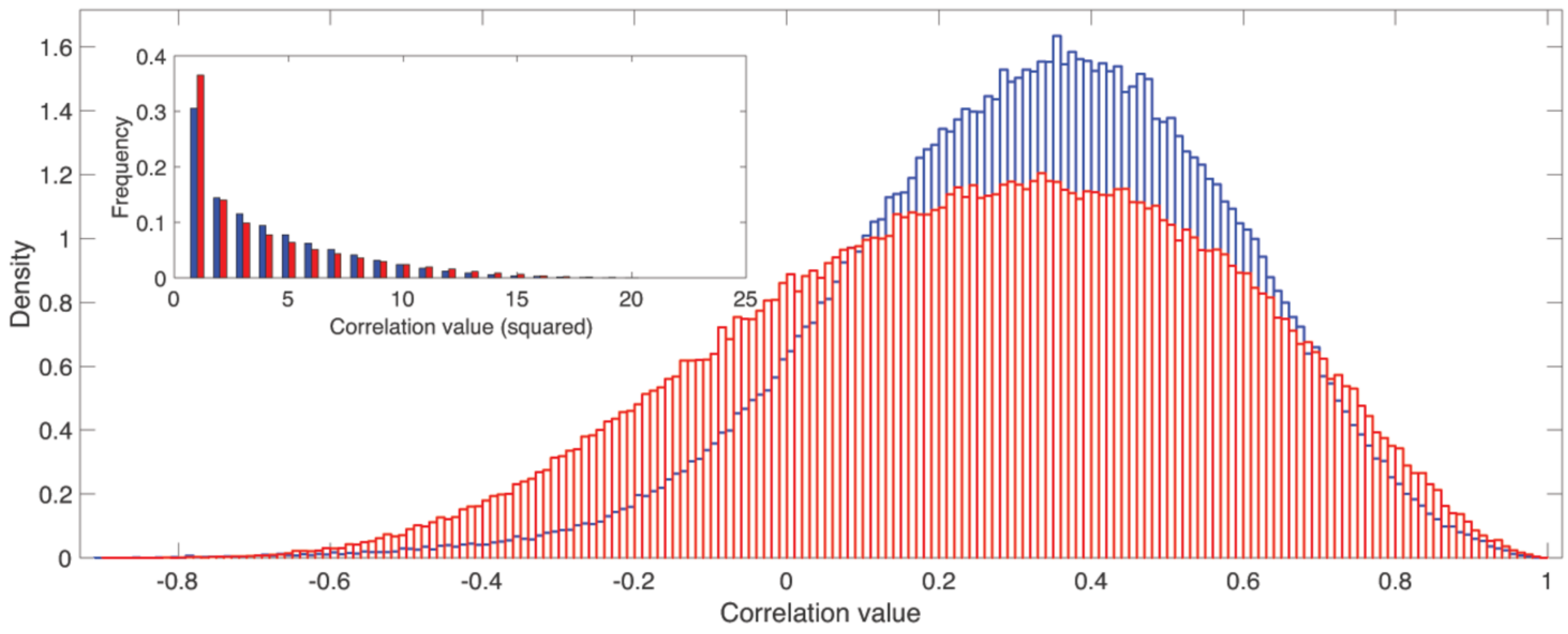
Density
1.6
1.4
1.2
1
0.8
0.6
0.4
0.2
0
-0.8
-0.6
-0.4
-0.2
0
0.2
0.4
0.6
0.8
1
Correlation value
Frequency
0.4
0.3
0.2
0.1
0
0
5
10
15
20
25
Correlation value (squared)

Figure 7

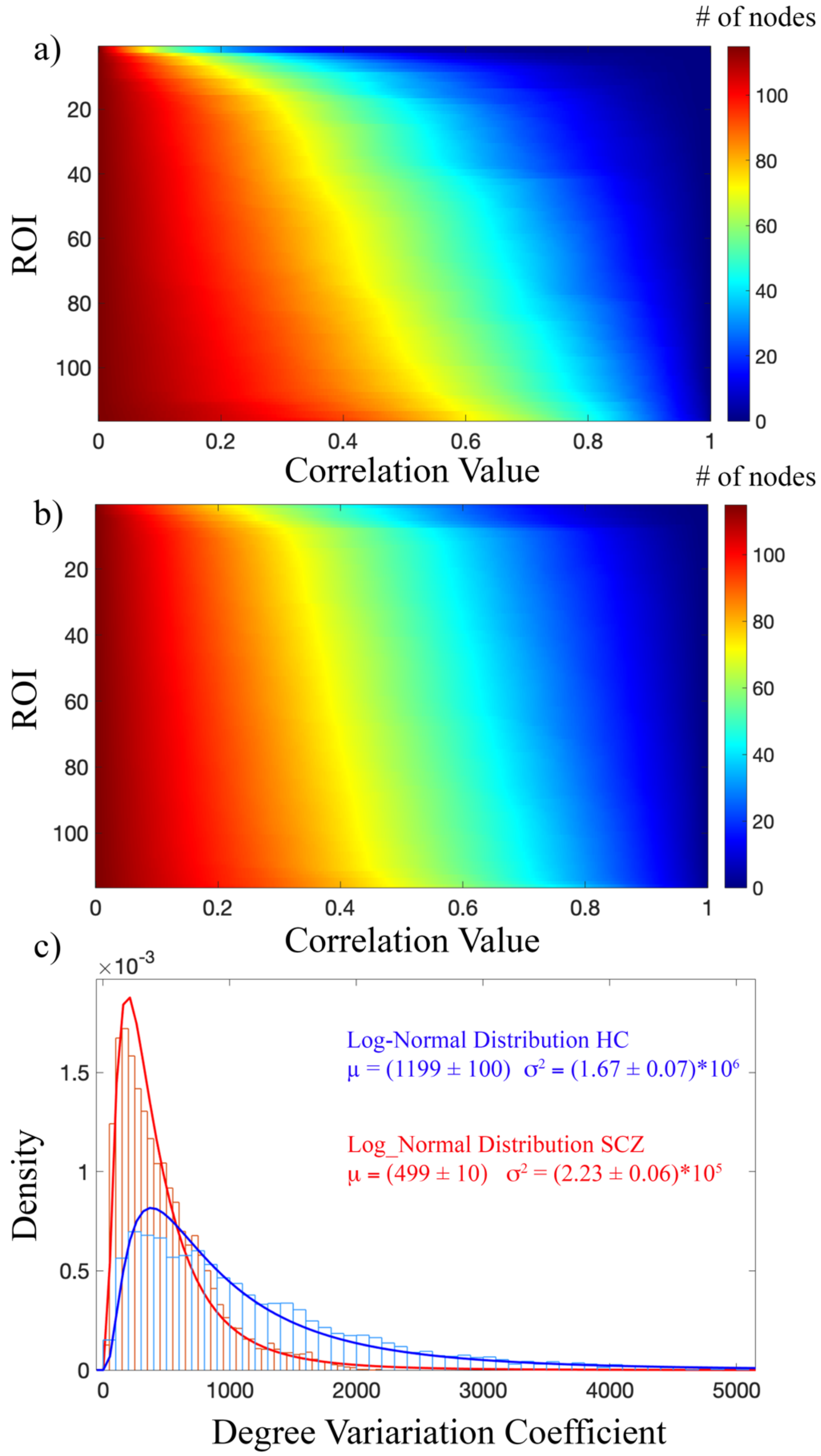

a)
# of nodes
ROI
Correlation Value
b)
# of nodes
ROI
Correlation Value
c)
Log-Normal Distribution HC
μ = (1199 ± 100) σ² = (1.67 ± 0.07)*10⁶
Log_Normal Distribution SCZ
μ = (499 ± 10) σ² = (2.23 ± 0.06)*10⁵
Density
Degree Variariation Coefficient

Figure 8

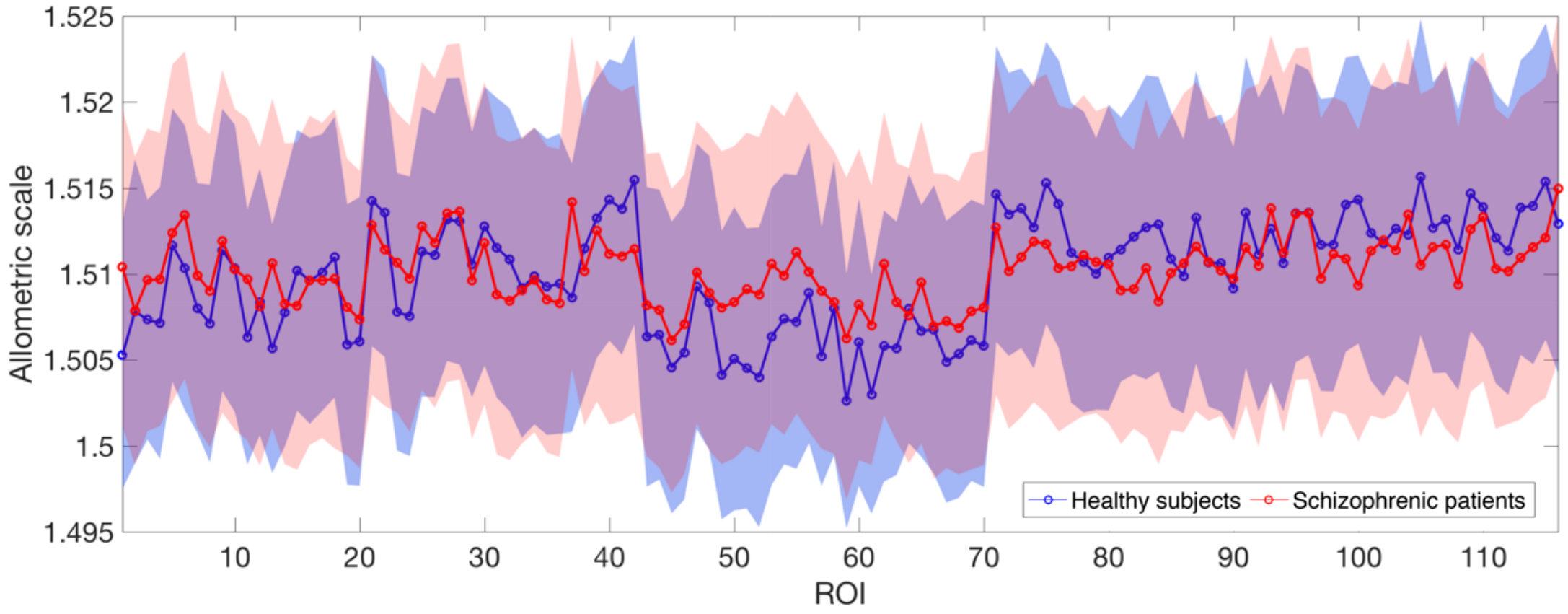
1.525
1.52
1.515
1.51
1.505
1.5
1.495
Allometric scale
10
20
30
40
50
60
70
80
90
100
110
ROI
Healthy subjects
Schizophrenic patients

Figure 9

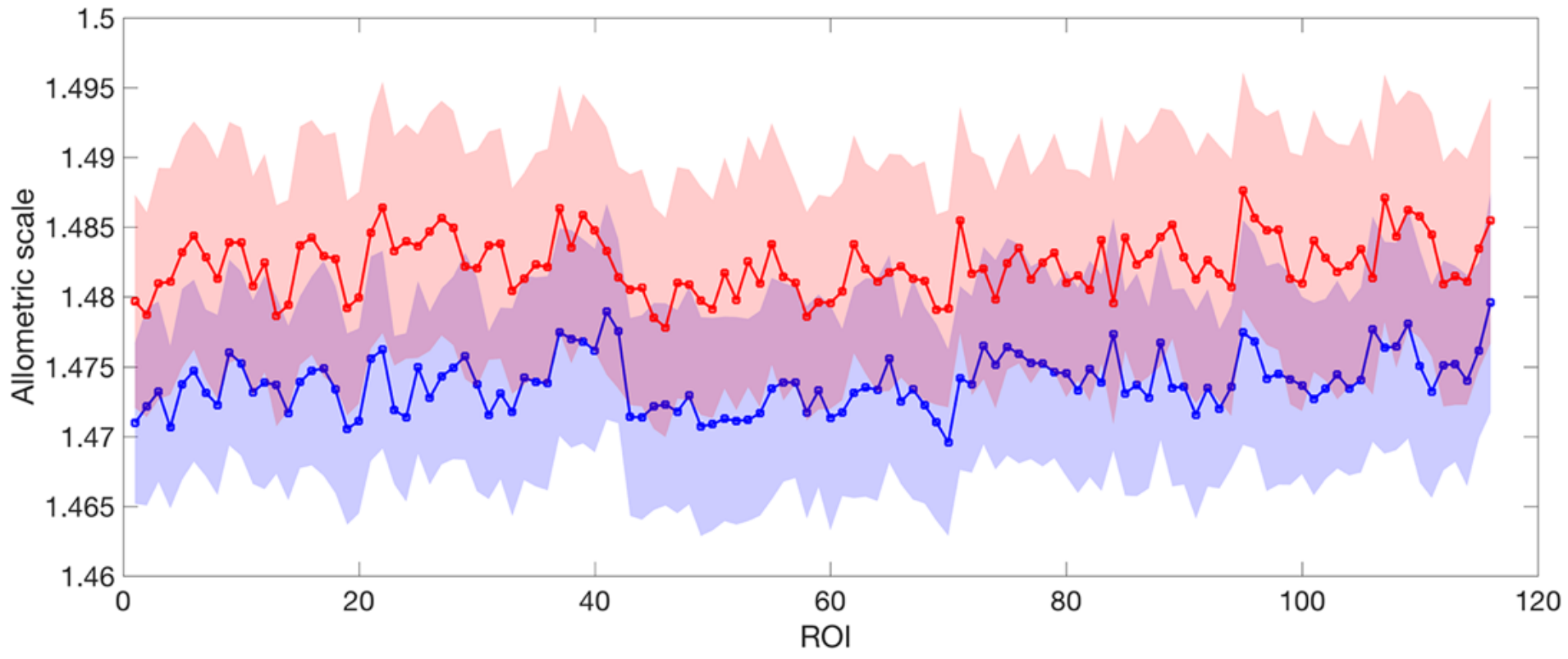

a)

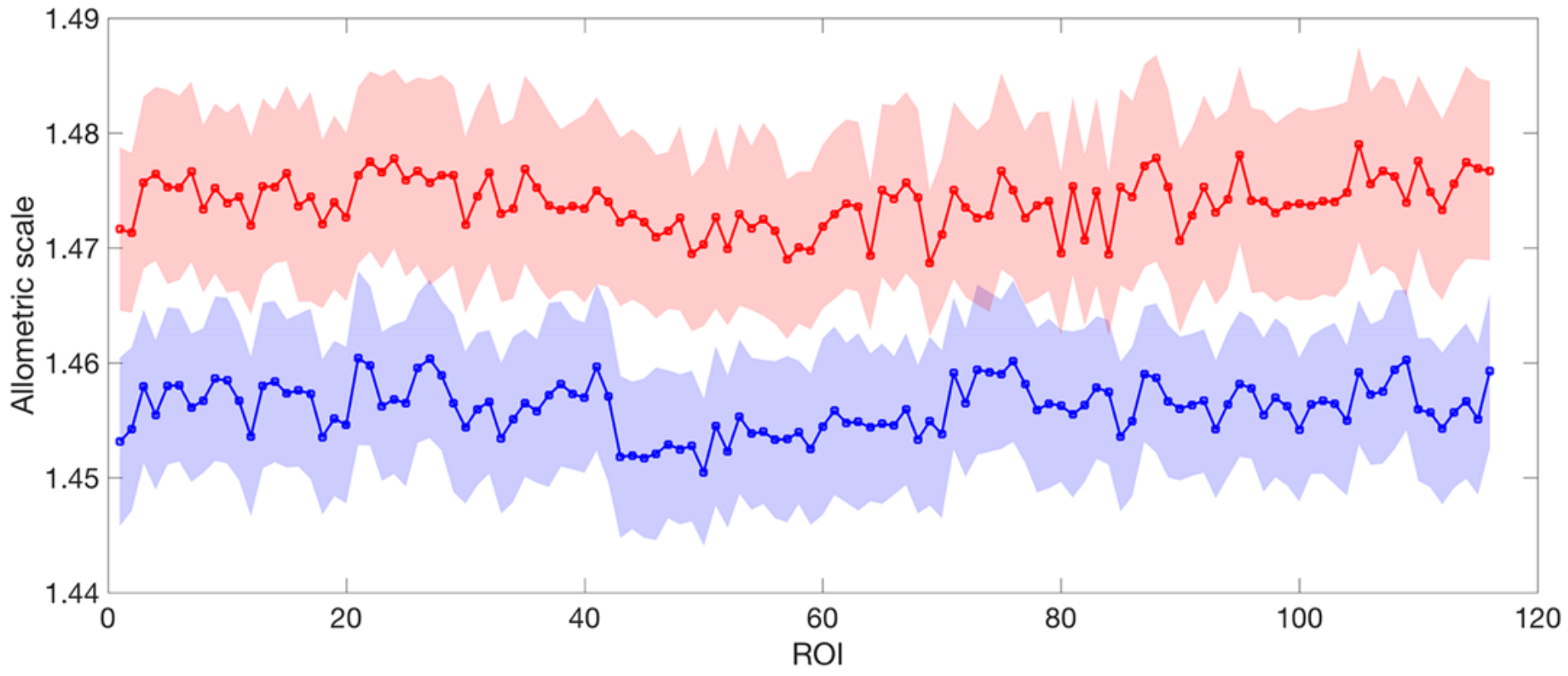

b)

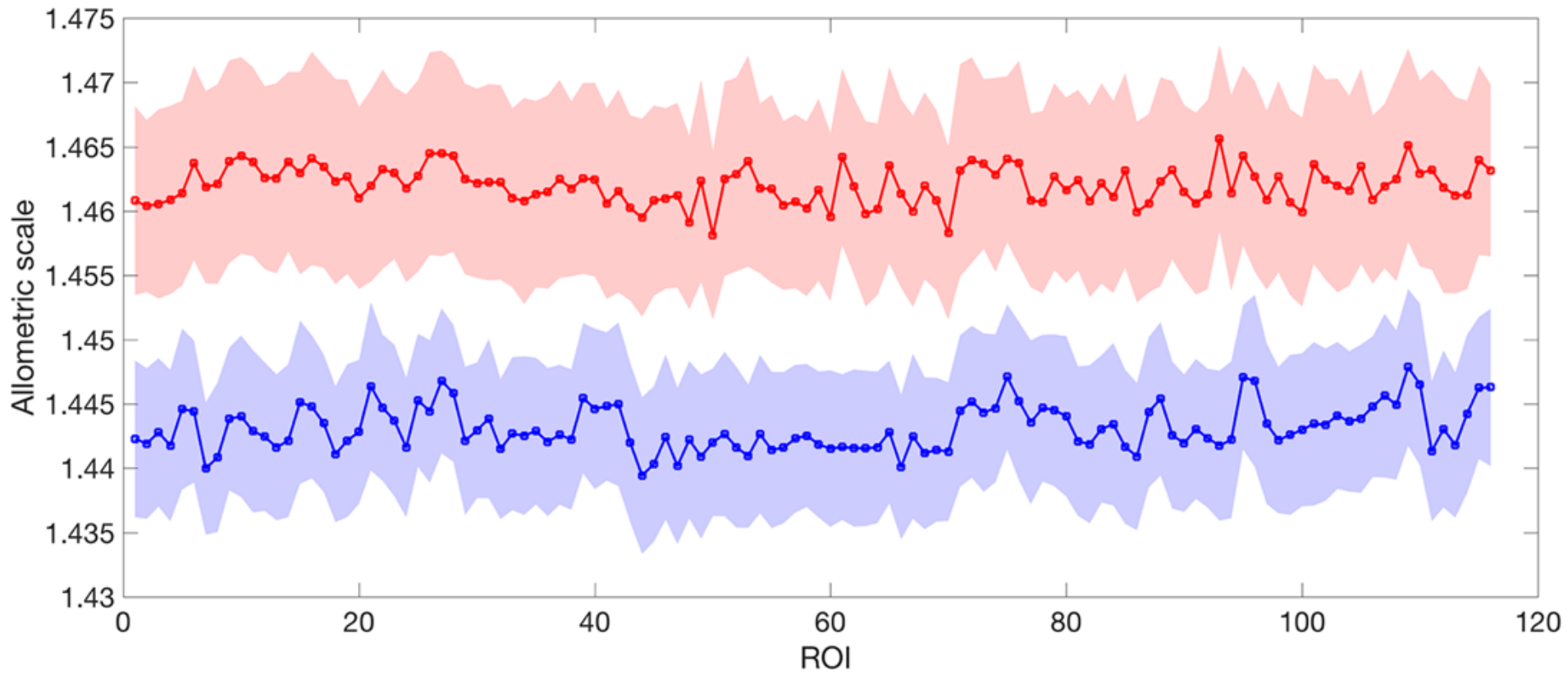

c)

Healthy subjects — Schizophrenic patients